\documentclass[a4paper,11pt]{article}
\usepackage{jheppub}
\usepackage{amsmath}
\usepackage{amsfonts}
\usepackage{amssymb}
\usepackage{textcomp}
\usepackage{graphicx}
\usepackage{bm}
\usepackage{color}
\usepackage{verbatim}
\usepackage{subcaption}

\usepackage{epsfig}

\usepackage{amsfonts,amsmath,amssymb}

\newcommand{\be}{\begin{equation}}
\newcommand{\ee}{\end{equation}}
\newcommand{\ba}{\begin{eqnarray}}
\newcommand{\ea}{\end{eqnarray}}
\newcommand{\ben}{\begin{enumerate}}
\newcommand{\een}{\end{enumerate}}

\newcommand{\p}{\partial}

\newcommand{\la}{\langle}
\newcommand{\ra}{\rangle}
\newcommand{\lr}{\leftrightarrow}

\newcommand{\rar}{\rightarrow}

\begin{document}

\preprint{NORDITA-2015-71}

\title{Phases of holographic d-wave superconductor}
\author{A. Krikun}
\affiliation{NORDITA \\
KTH Royal Institute of Technology and Stockholm University \\
Roslagstullsbacken 23, SE-106 91 Stockholm, Sweden. }

\affiliation{Institute for Theoretical and Experimental Physics (ITEP) \\
B. Cheryomushkinskaya 25, 117218 Moscow, Russia}

\emailAdd{krikun@nordita.org}

\abstract{We study different phases in the holographic model of d-wave superconductor. These are described by solutions to the classical equations of motion found in different ansatze. Apart from the known homogeneous d-wave superconducting phase we find three new solutions. Two of them represent two distinct families of the spatially modulated solutions, which realize the charge density wave phases in the dual theory. The third one is the new homogeneous phase with nonzero anapole moment. These phases are relevant to the physics of  cuprate high-Tc superconductor in pseudogap region.

While the d-wave phase preserves translation, parity and time reversal symmetry, the striped phases break translations spontaneously. Parity and time-reversal are preserved when combined with discrete half-periodic shift of the wave. In anapole phase translation symmetry is preserved, but parity and time reversal are spontaneously broken. All of the considered solutions brake the global $U(1)$.

Thermodynamical treatment shows that in the simplest d-wave model the anapole phase is always preferred, while the stripe phases realize the continuous transition in solution space between the normal phase and two homogeneous condensed phases.
}


\notoc

\maketitle

\section{Introduction}
The large variety of different phases, which is observed in the cuprate superconductors is a distinctive feature of high-T$_c$ superconductivity and a big puzzle in modern condensed matter theory. The superconducting dome is surrounded by the normal metal, strange metal and pseudogap phases. While the normal metal is described by the Fermi liquid theory and is relatively well understood, the other two phases haven't got yet a rigorous theoretical description. The strange metal is believed to be a normal phase of the strongly coupled medium, which may not allow for a quasiparticle description and thus exhibits the features, like linear in temperature resistivity, which can not be understood in the framework of the standard Fermi-liquid approach. The pseudogap may be though of as a phase where some symmetries of this strongly coupled medium are broken by a bunch of competing order parameters, leading to the very controversial phenomenological features, which are not well understood even from the experimental point of view. 
In the last few years a considerable experimental evidence and theoretical understanding has been accumulated concerning the presence of the charge density wave in the pseudogap region of the cuprate phase diagram, which breaks translational invariance in the material by introducing a striped superstructure, which may or may not be commensurate with the lattice spacing \cite{Gruner:1988zz, hashimoto2010particle, chang2012direct, efetov2013pseudogap}. From the other hand, there were claims that in the pseudogap one can observe a so called loop current order \cite{varma2006theory, kaminski2002spontaneous, varma2010high}, the mesoscopic ordering of the molecular currents, which is described by nonzero in-plane time-reversal odd polar vector, so called ``anapole moment''. The claims for other exotic orderings are often arising either in theoretical or in experimental literature \cite{chakravarty2001hidden}. In general the clear understanding on what is actually happening in the pseudogap region is missing so far. 
 
The reason why these unconventional phases of cuprates remain an unsolved problem of physics is the fact that the underlying system is strongly coupled. In the case of strange metal this leads to the absence of quasiparticles and in the case of pseudogap it substantially mixes the various order parameters. The modern tool, which allows one to treat the strongly coupled theories in a relatively controlled fashion, is the holographic duality. First formulated as a duality between strongly coupled supersymmetric N=4 Yang-Mills theory and string theory in a curved background, this approach can be generalized to the other strongly coupled systems. Even though there is no rigorous proof of the duality even in a simple supersymmetric case, the holographic approach provides a useful algorithm for construction of phenomenological models which can point out unexpected links between the features of the system which could not be related via the other approaches. Nowadays there is a bursting activity in applying the 
holographic tools to the study of the strange metals \cite{Hartnoll:2014lpa, Hartnoll:2015sea, Davison:2013txa, Davison:2014lua, Davison:2015bea, Lucas:2015vna}. This paper is devoted to the holographic study of the possible phases and order parameters, which can arise in a strongly coupled medium like cuprate high-T$_c$ superconductor in a pseudogap region.

The paper is organized as follows. In Section \ref{Sec:d-wave} we introduce the holographic model of d-wave superconductor and describe a d-wave superconducting phase, which is present there, as well as its spatially modulated unstable modes. In Section \ref{stripes} we obtain the spatially modulated solutions in this model and show that they include charge density wave as seen from the dual field theory perspective. In this study we also find a strong indication of the presence of another homogeneous phase of the model, which has not been observed before in the literature. We discuss the thermodynamics and special features of this new phase in Section \ref{Sec:anapole}. Conclusion is given in Section \ref{conclusion}. The single Appendix is devoted to the details of the numerical calculation performed in Section \ref{stripes}.

\section{\label{Sec:d-wave}Holographic model of D-wave superconductor}

The holographic model of d-wave superconductor was introduced in \cite{Chen:2010mk, Benini:2010pr, Benini:2010qc}. The top-down approach to this problem was discussed in \cite{Kim:2013oba}. The main ingredient of the model is a charged massive spin-2 field which is dual to the d-wave order parameter. The consistent action for such a field on the curved background can be written down in the following way 
\begin{align}
\label{action}
\mathcal{L} =& -|D_\rho \phi_{\mu \nu}|^2 + 2|D_\mu \phi^{\mu \nu}|^2 + |D_\mu \phi|^2 - [D_\mu \phi^{*\mu \nu} D_\nu \phi + c.c.] - m^2 (|\phi_{\mu \nu}|^2 - |\phi|^2) \\
\notag
& + 2 R_{\mu \nu \rho \lambda} \phi^{*\mu \rho} \phi^{\nu \lambda} - \frac{1}{d+1} R|\phi|^2 - iqF_{\mu \nu} \phi^{* \mu \lambda} \phi_{\lambda}^{\nu} - \frac{1}{4}F_{\mu \nu} F^{\mu \nu},
\end{align}
where $\phi_{\mu \nu}$ is a complex symmetric tensor field, $\phi=\phi_{\mu}^\mu$, $\phi_\mu = D^{\nu} \phi_{\mu \nu}$ and covariant derivative acts on it as $D_\mu = \nabla_\mu - i q A_\mu$. In \cite{Benini:2010pr} it was discussed, that in order to describe a proper number of the degrees of freedom and avoid ghosts in the spectrum, one needs to  neglect the backreaction of the matter fields on the metric. This can be done in a consistent way by taking the limit of infinitely large gauge charge $q\rar \infty$. We should stress here that neglecting gravitational backreaction doesn't mean that the theory becomes linear, as the self interaction of the matter field is not suppressed and leads to an interesting nonlinear dynamics. Hence in the following we are considering the dynamics of the tensor and gauge fields on top of the static gravitational background. In order to describe the system at finite temperature one chooses AdS/Schwarzschild black hole as a background metric:\footnote{Keeping in mind that cuprates have 
layered structure, i.
e. the system we aim to describe is quasi 2+1 dimensional, the number of dimensions in the holographic model is 3+1.}
\begin{equation}
\label{metric}
d s^2 = \frac{L_0^2}{\tilde{z}^2} \left(- f(z) dt^2 + f(z)^{-1} dz^2 + dx^2 + dy^2 \right), \qquad f(z) = 1 - \frac{z^3}{z_h^3}.
\end{equation}
and the radius of the horizon is related to the temperature in the dual system $T = \frac{3}{4 \pi} \frac{1}{z_h}$. In what follows we will rescale the curvature radius of the space to unity: $L_0 = 1$.

In \cite{Benini:2010pr} the specific ansatz for the tensor field was considered. It included only the spatial components of the tensor field $\phi_{xy} = \phi_{yx} \neq0$, which are dual to the charged operator with $d_{xy}$ spherical symmetry, and temporal component of the gauge field dual to the charge density $\rho$ and chemical potential $\mu$. This ansatz provides a consistent truncation to the equations of motion. At high temperatures the only existing solution is the empty Reissner-Nordstr\"om black hole with
\begin{equation}
\label{RN}
A^{NS}_t = \mu \left(1 - \frac{z}{z_h} \right), \qquad \psi_{\mu \nu} = 0
\end{equation}
Below the certain critical temperature $T_c$, defined in units of $\mu$, the nontrivial solution to the equations of motion arises which is related by duality to the emergence of nonzero vacuum expectation value (VEV) of the order parameter. Hence it was shown that the model describes a transition from the normal phase to the superconducting (SC) phase with $d_{xy}$ gap, similar to the transition observed in the holographic model of s-wave superconductor \cite{Hartnoll:2008kx, Hartnoll:2008vx}. 

\begin{figure}[ht]
\centering
\includegraphics[width=0.6 \linewidth]{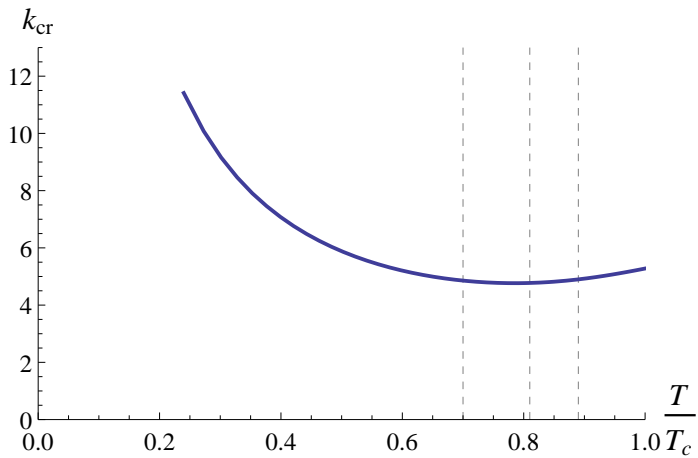} 
\caption{\label{kcs} The values of momenta of the unstable mode (measured in units of $z_h^{-1}$) in the perturbative spectrum of d-wave superconducting phase. See \cite{Krikun:2013iha} for details. Dashed lines show the temperatures used for plots in Fig.\,\ref{Omega_plots}. }
\end{figure}

Contrary to the s-wave model, though, the d-wave holographic action (\ref{action}) contains several terms which contribute to the mixing between different components of the tensor field. And these terms are proportional to the spatial momentum of the corresponding modes. This observation motivated our earlier study of the possible spatially modulated instabilities of the condensed phase in the spirit of \cite{Donos:2011bh, Donos:2013gda, Donos:2013wia, Rozali:2012es, Nakamura:2009tf}. The mixing terms proportional to momentum  are responsible to the spontaneous translational symmetry breaking in these models. Indeed, in \cite{Krikun:2013iha} we found that the perturbative spectrum of the condensed phase of holographic d-wave superconductor has a spatially modulated unstable mode with the specific wave vector $k_c$, which signals the onset of the instability leading to the formation of the new striped phase. Moreover, the instability was found at all temperatures, the fact which means that the condensed SC phase is always unstable in the model under consideration, see Fig.\,\ref{kcs}.

The perturbative study of the spectrum performed in \cite{Krikun:2013iha} can show the presence of instability, but can not describe the phase, which is the endpoint of the transition. In order to study the new phase one need to solve full nonlinear system of equations of motion.

\section{\label{stripes}Striped phases}
The unstable mode observed in \cite{Krikun:2013iha} includes the fluctuations of several components of the tensor field as well as the fluctuation of the temporal component of the gauge field dual to the charge density. The instability is modulated only along either $x$- or $y$-axis (where the axes are defined in such a way, that it is component $\phi_{xy}$ which condenses in the superconducting phase). If one considers modulation along $y$-axis with the wave vector $k_y$, the components involved in the unstable mode are $\phi_{tx}, \phi_{zx}, \phi_{xy}$ and $A_t$. The symmetry of the model requires also the existence of the similar mode in $x$-direction which is obtained from the previous one by simple exchange $x \lr y$. This form of the unstable mode suggests the following ansatz for the solution to the \textit{nonlinear} equations of motion  
\begin{gather}
\phi_{\mu \nu} = \frac{1}{2 z^2} \frac{L}{q} \left(\begin{matrix}
        0 & i \psi_{tx}(y,z) &0 & 0 \\
        i \psi_{tx}(y,z) & 0 & \psi_{xy}(y,z) & \psi_{xz}(y,z) \\
        0 & \psi_{xy}(y,z) & 0 & 0 \\
        0 & \psi_{xz}(y,z) & 0 & 0        
       \end{matrix}\right), \\
       \notag
A_\mu = \left(\begin{matrix}
          A_t(y,z) & 0 & 0 & 0
         \end{matrix}\right),
\end{gather}
where $\psi_{\mu \nu}$ are the real functions and the overall factor coincides with the one used in \cite{Benini:2010pr}. By plugging this ansatz into the equations of motion, which follow from the action (\ref{action}), one can check that this is indeed a consistent truncation of the model at nonlinear level. The only nontrivial equations of motion in this ansatz are
\begin{align}
\label{EOMs}
E_{xy}: \quad  &\p_z^2 \psi _{xy} +\left(\frac{f'(z)}{f(z)}-\frac{2}{z}\right) \p_z \psi_{xy} + \psi _{xy}  \left(\frac{A_t^2}{f(z)^2}-\frac{L^2 m^2}{z^2 f(z)}\right) \\
\notag
&\qquad  - \p_z \p_y \psi_{zx}   +\left(\frac{2}{z}-\frac{f'(z)}{f(z)}\right) \p_y \psi_{zx} +\frac{\p_y A_t \ \psi _{tx}}{2 f(z)^2}+\frac{A_t \ \p_y \psi_{tx} }{f(z)^2}  = 0,\\
\notag
E_{tx}: \quad & \p_z^2 \psi_{ tx} -\frac{2}{z} \p_z \psi_{tx} +\frac{\p_y^2 \psi_{ tx}}{f(z)} -\frac{  \left(z^2 f''(z)-2 z f'(z)+2 L^2 m^2\right)}{2 z^2 f(z)} \psi_{ tx} \\
\notag
&\qquad + \frac{\p_y A_t \ \psi_{ xy} }{2 f(z)}+\frac{A_t \ \p_y \psi_{ xy} }{f(z)} + A_t \ \p_z \psi_{ zx} +\left(\frac{1}{2} \p_z A_t -\frac{2 A_t }{z}\right) \psi_{ zx} =0,\\
\notag
E_{zx}: \quad & \p_y^2 \psi_{zx} -\p_z \p_y \psi_{xy} + \psi_{zx}  \left(\frac{A_t^2}{f(z)}-\frac{f''(z)}{2}+\frac{f'(z)}{z}-\frac{ L^2 m^2}{z^2}\right) \\
\notag
& \qquad +\frac{\p_z A_t \psi_{ tx} }{2 f(z)}+\frac{A_t  \p_z \psi_{ tx} }{f(z)}  =0, \\
\notag
E_t: \quad & \p_z^2 A_t +\frac{\p_y^2 A_t }{f(z)} -\frac{A_t  \psi_{ xy}^2}{z^2 f(z)}-\frac{A_t  \psi_{zx}^2}{z^2} \\
\notag
& \qquad -\frac{\p_y \psi_{tx} \psi_{ xy} }{2 z^2 f(z)}+\psi_{ tx}  \left(\frac{\p_y \psi_{ xy}}{2 z^2 f(z)}-\frac{\psi_{zx} }{z^3}+\frac{\p_z \psi_{ zx}}{2 z^2}\right)-\frac{\p_z \psi_{ tx} \psi_{ zx} }{2 z^2}=0.
\end{align}
On top of that the expression $D^{\mu} E_{\mu \nu} = 0$ leads to a nontrivial constraint:
\begin{align}
\label{Constr:striped}
C_x: \ &   \p_y \psi_{ xy}  + f(z) \p_z \psi_{ zx}  - \frac{z^2 \psi_{tx}}{2 m^2} \left( \p_z^2 A_t  + \frac{\p_y^2 A_t }{f(z)} \right) -\frac{A_t \psi _{tx} }{f(z)}+\left(f'(z)-\frac{4}{z} f(z)\right)\psi_{ zx}   \\
\notag
& - \frac{3 z^2}{2 m^2 f(z)} \p_y A_t \left( A_t  \psi _{ xy}+  \p_y \psi_{ tx} \right) + \frac{z^2}{2 m^2 } \p_z A_t \left( \frac{2}{z} \psi_{ tx} -3 A_t  \psi_{ zx} - 3  \p_z \psi_{ tx}\right)=0.
\end{align}

We study the solution to these equations numerically. We expect the solution to be periodic in $y$-direction with a wavelength $\lambda_y=\frac{2 \pi}{k_y}$, thus we can choose a finite domain $y \in (0,\lambda_y), z \in (0,z_h)$ in which the equations of motion have to be solved. Periodic boundary conditions are imposed in $y$-direction. The boundary conditions in $z$-direction are specified by means of usual holographic prescriptions.

Near the $z=0$ boundary the fields can be expanded in terms of normalizable and non-normalizable modes:
\begin{align}
\label{asympt}
\psi_{\mu \nu} (y,z)&= \psi^{(0)}_{\mu \nu}(y)  \left( z^{\frac{3}{2}-\frac{1}{2}\sqrt{9+4 m^2}} + \dots \right)+ \psi^{(1)}_{\mu \nu}(y) \left( z^{\frac{3}{2}+\frac{1}{2}\sqrt{9+4 m^2}} + \dots \right), \qquad \mu,\nu \neq z \\
\notag
\psi_{\mu z} (y,z)&= \psi^{(0)}_{\mu z}(y)  \left( z^{\frac{5}{2}-\frac{1}{2}\sqrt{9+4 m^2}} + \dots \right)+ \psi^{(1)}_{\mu z}(y) \left( z^{\frac{5}{2}+\frac{1}{2}\sqrt{9+4 m^2}} + \dots \right), \qquad  \mu \neq z \\
\notag
A_\mu(y,z) &= A_\mu^{(0)}(y) + A_\mu^{(1)}(y) z + \dots
\end{align}
The coefficients for leading modes are equal to the sources of corresponding operators in the dual theory, while the coefficients for subleading modes are related to the vacuum expectation values.\footnote{After plugging in these expressions to the equations (\ref{EOMs}) one can notice, that the source term for $\psi_{xz}$ is related to the sources of the other fields. Thus there is no independent source term for the $\psi_{xz}$ component. It is expected indeed, because this component is not dual to any operator in the boundary theory.} The only operator source, which is present in the problem, is a chemical potential $\mu$, so the boundary conditions at $z=0$ are 
\begin{align}
\label{bcB}
 A_t^{(0)}(y) = \mu, \qquad \psi^{(0)}_{\mu \nu}(y) = 0.
\end{align}

The boundary conditions at the black hole horizon are specified by the usual requirement that only the modes infalling under horizon are present \cite{Son:2002sd}. In case of zero frequency this just means that the fields are regular at $z=z_h$.\footnote{For $A_t$ and $\psi_{tx}$ regularity requires the functions to vanish at the horizon linearly in $f(z)$. In our numerical procedure we solve for the nonvanishing functions $\hat{A}_t = f(z)^{-1} A_t$ and $\hat{\psi}_{tx} = f(z)^{-1} \psi_{tx}$. See also (\ref{rescale}) in the Appendix} In order to impose such regularity conditions we expand the equations of motion near the horizon. The leading terms in $z$ expansion form a system of 4 nonsingular differential equations (\ref{boundary_eq}), which can be seen as a system of ordinary differential equations along the $y$-coordinate with the boundary values of the fields and their first $z-$derivatives being unknown functions.\footnote{One should pay attention to the fact that the equation $E_{xy}$ in (\ref{EOMs}) does not include the second derivative on $y$. Hence in the boundary expansion only 3 equations of 4 are of the second order and the boundary value problem along $y$-axis is not well defined. In our calculation it was useful to avoid this complication by considering the constraint (\ref{Constr:striped}) which relates the derivatives $\p_y \psi_{tx}$ and $\p_z \psi_{zx}$. By using this constraint one can re-express the term $\p_z \p_y \psi_{zx}$ in $E_{xy}$ via $\p_y^2 \psi_{xy}$ and obtain the second order equation for $\psi_{xy}$.} These relations provide us with the generalized boundary conditions at the horizon. 

Given these boundary conditions we find solutions to the system of 2D PDEs (\ref{EOMs}) by means of the finite difference derivative method, approximating the derivatives pseudospectrally and using Newton-Raphson procedure in order to solve the resulting system of algebraic equations (see details in the Appendix \ref{App:Numerics}). 

For each numerical solution we extract its thermodynamical (TD) potential. By holographic duality it is related to the Euclidean action, evaluated on the given solution. As the equations of motion are satisfied, the action reduces to the boundary term \cite{Bianchi:2001kw, Skenderis:2002wp}. According to the condition (\ref{bcB}) the only contribution to this boundary term comes from the kinetic action of the gauge field which is proportional to $\mu$. Hence the expression for the mean density of thermodynamic potential is
\begin{align}
\label{Omega}
\omega = - \frac{1}{q^2}  \int \limits_{0}^{\lambda_y} \frac{dy}{\lambda_{y}}  \ \mu \rho(y),
\end{align}
where $\rho(y) = - A^{(1)}_t (y)$ is the expectation value for the local charge density in the boundary theory. It is useful to note here, that the thermodynamic potential of the solutions under consideration is of order $q^{-2}$. If one would be able to consider the backreaction of matter fields on the geometry, the perturbation of the metric would be $\delta g = O(q^{-2})$ and the corresponding contribution of gravity to the thermodynamic potential would be $O(\delta g^2) \sim O(q^{-4})$ (the term $O(\delta g)$ vanishes due to the equations of motion) and is subleading in the $q\rar \infty$ regime. Hence it does actually make sense to study the thermodynamics of the matter fields even though the backreaction of the gravity can not be taken into account in the present setting.

\begin{figure}[ht]
\centering
\begin{subfigure}[b]{1\textwidth}
 \includegraphics[width=0.32\linewidth]{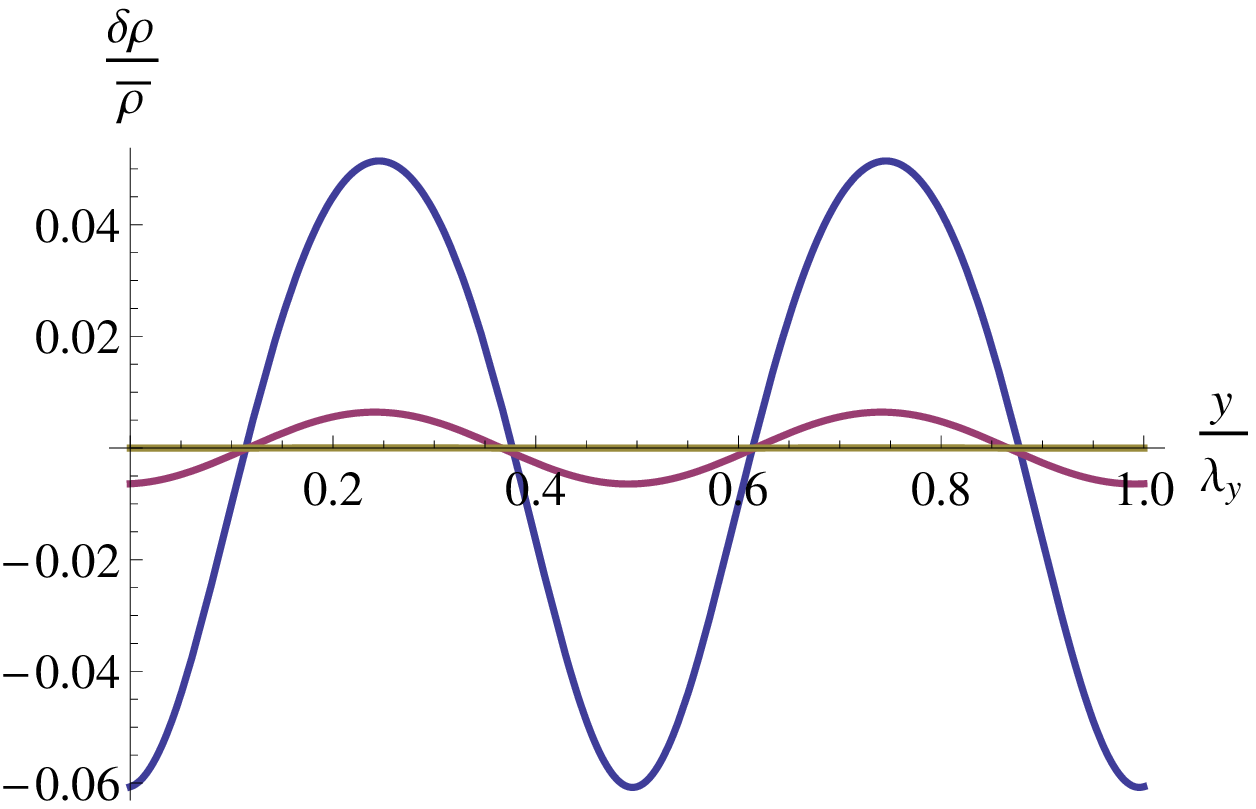} 
 \includegraphics[width=0.32\linewidth]{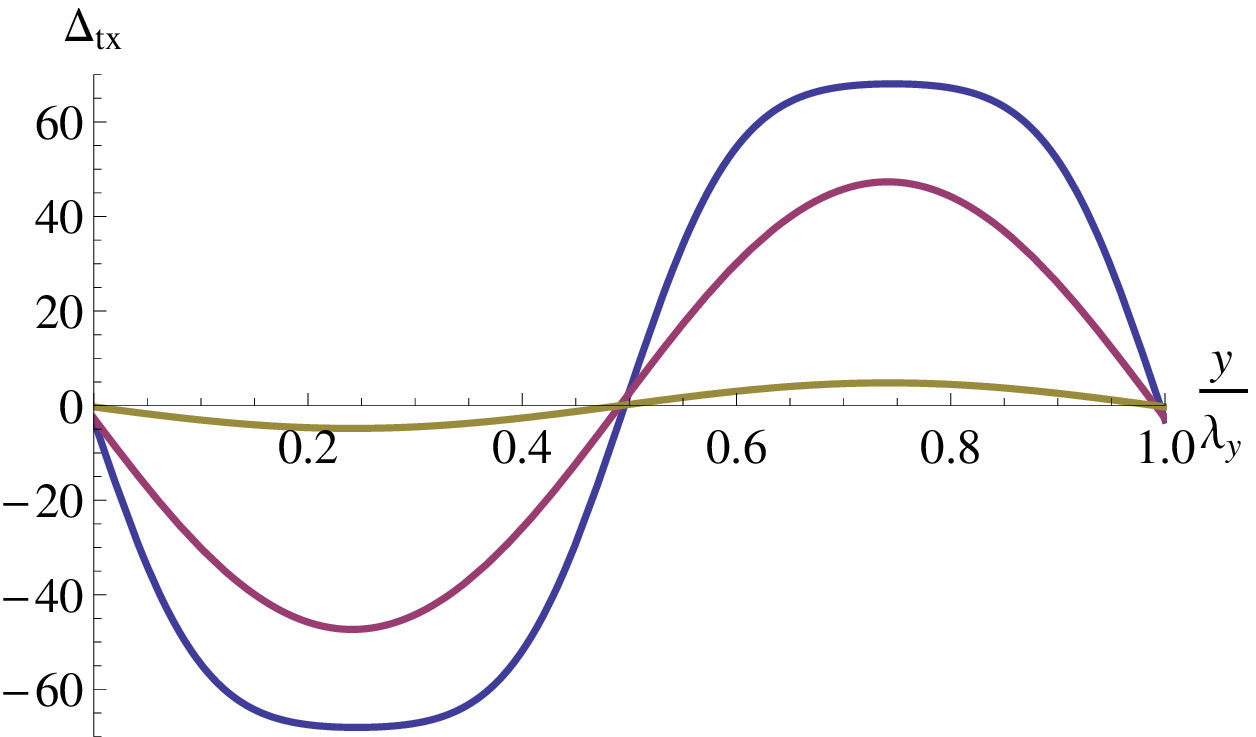} 
 \includegraphics[width=0.32 \linewidth]{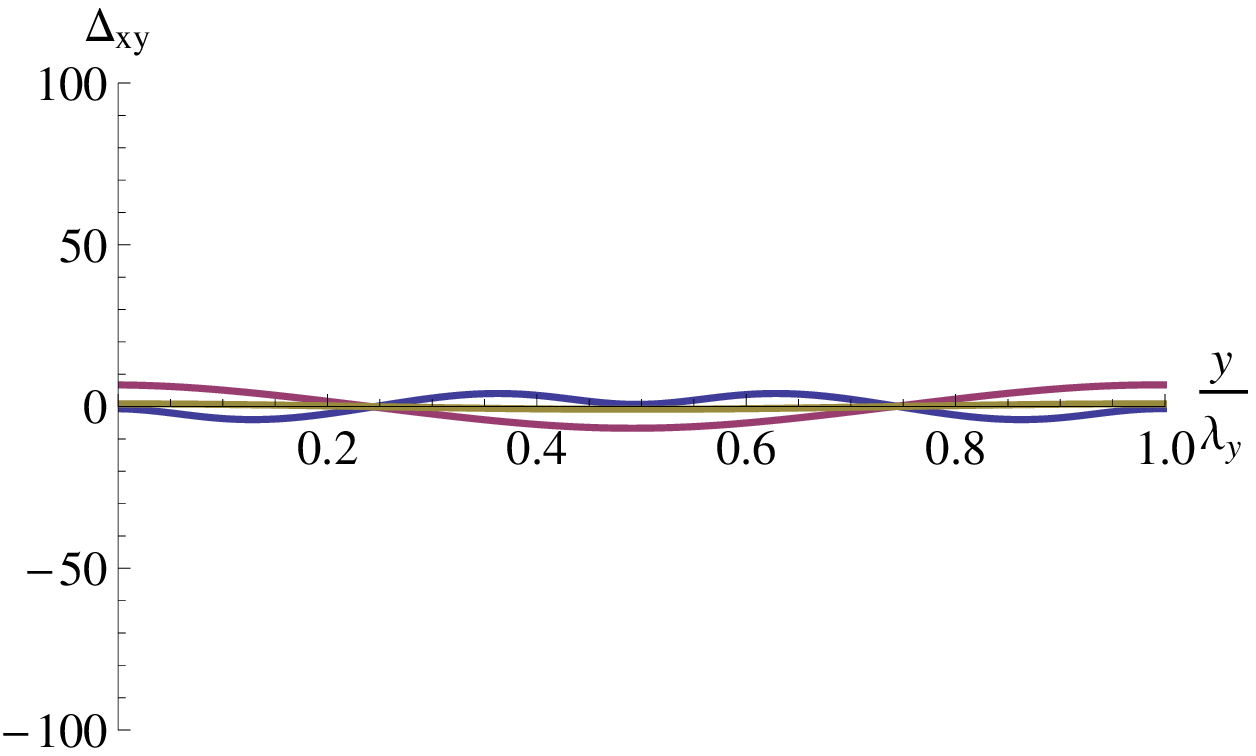} 
 \caption{\label{Type_A} Type A}
\end{subfigure} \\
\begin{subfigure}[b]{1\textwidth}
 \includegraphics[width=0.32\linewidth]{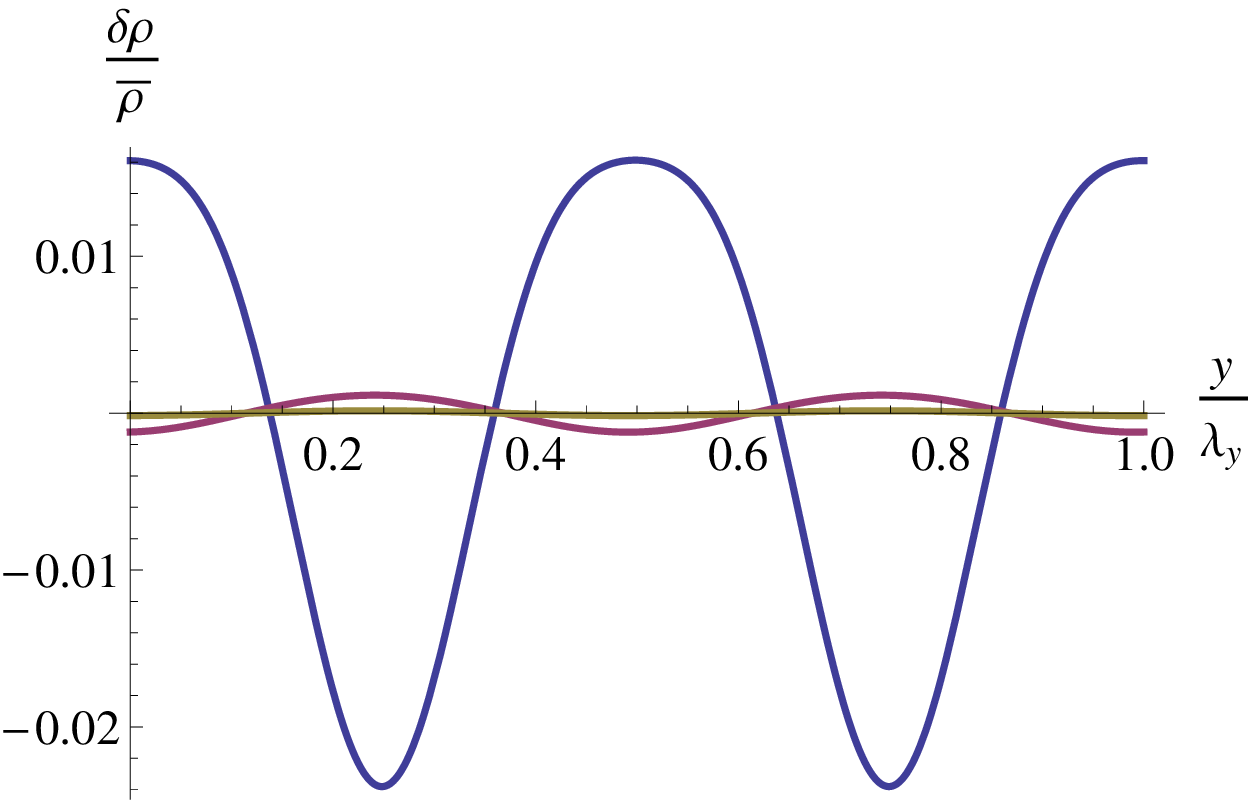} 
 \includegraphics[width=0.32\linewidth]{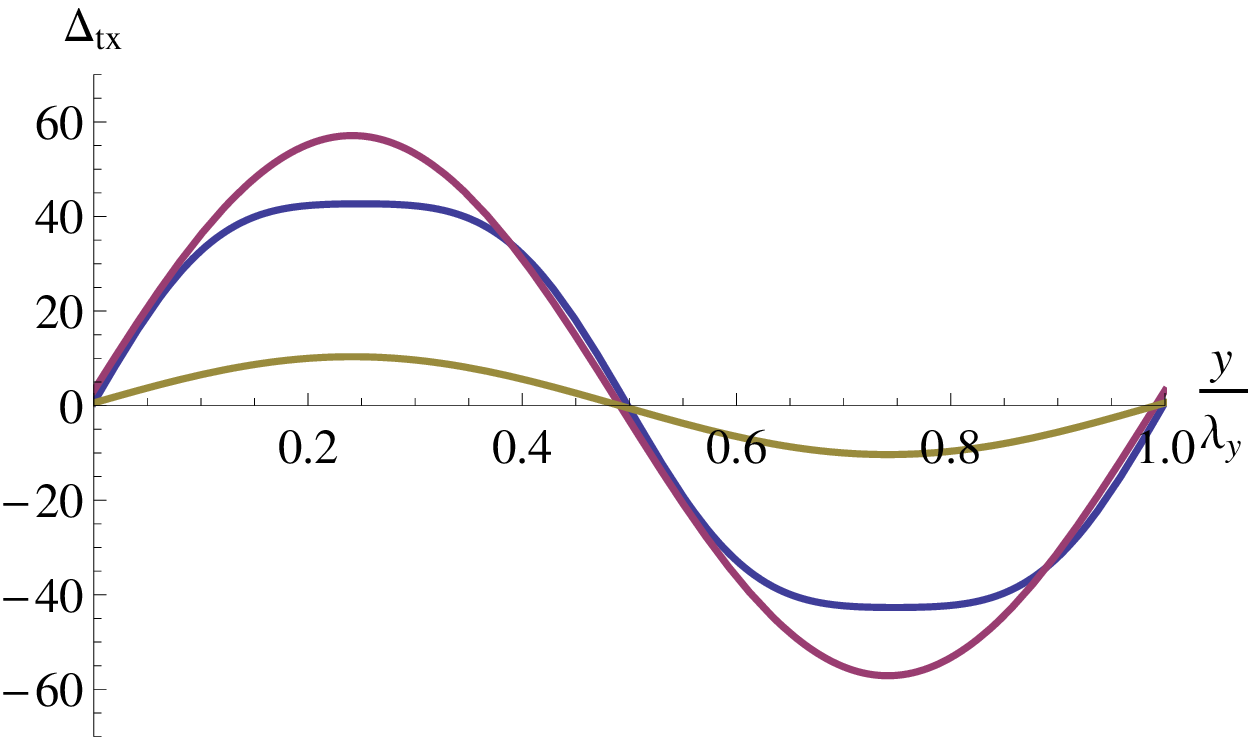} 
 \includegraphics[width=0.32 \linewidth]{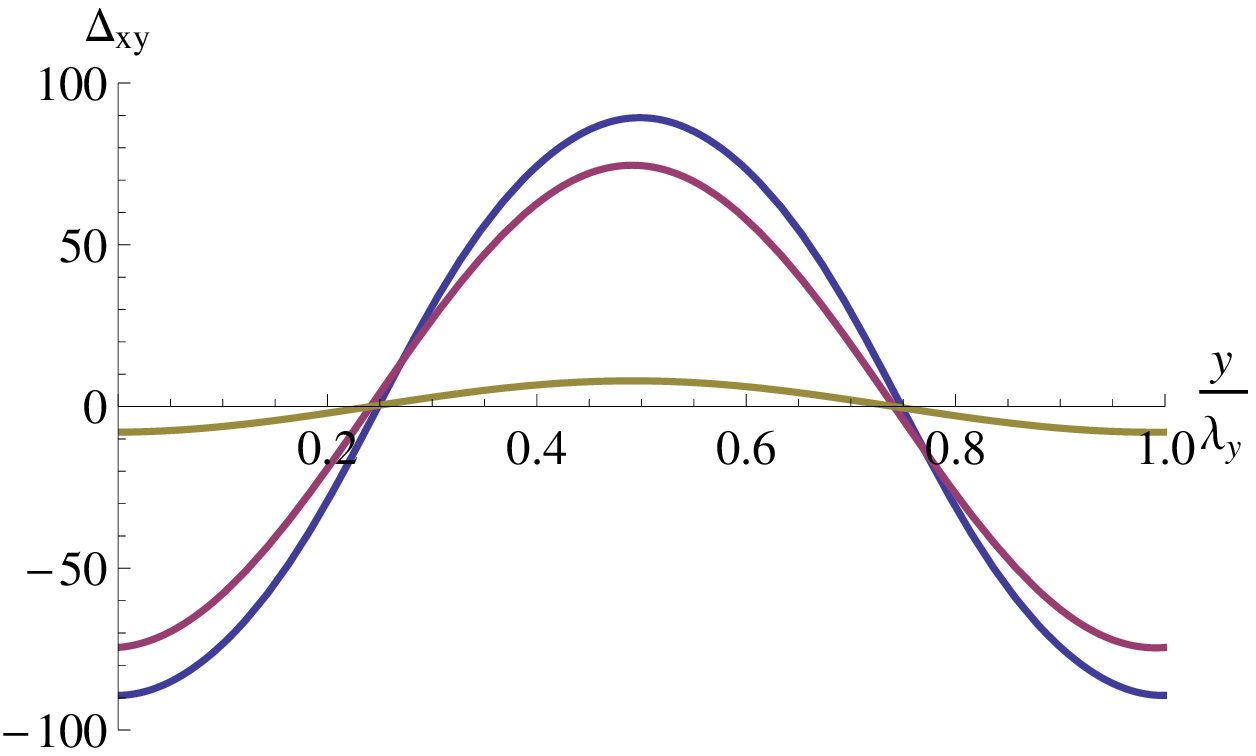} 
 \caption{\label{Type_B} Type B}
\end{subfigure}
\caption{\label{Stripes}Boundary data for the two families of striped solutions at $T\approx 0.8 \, T_c$ with the wave vectors $k_y z_h = 2, \, 4.8$ and $7.6$ for Type A and $k_y z_h = 1.9, \, 3.1$ and $5.2$ for Type B. The values for tensor condensates $\Delta$ are measured in units of $z_h$. The scales of plots for Type A and Type B is intentionally kept the same.}
\end{figure}

In this section we restrict ourselves to the study of the model with particular value of the tensor field mass $m^2 = 4$ which corresponds to the scaling dimension of the dual operator $[\Delta_{\mu \nu}] = 4$. In the following we will also measure the temperature in relation to the critical temperature of the d-wave superconductor phase transition. For chosen mass this temperature is \cite{Benini:2010pr}
\begin{equation}
T_c \approx  \frac{3}{4 \pi} \frac{1}{11.29} q \mu.
\end{equation}
We were able to find \textit{two distinct families} of striped solutions at given set of temperatures and wavelengths $\lambda_y$. We will denote them as Type A and Type B. After extracting the local values of the charge density from the asymptotic behavior of the solutions (\ref{asympt}), we find that in both families they are spatially modulated, see Fig.\,\ref{Stripes}. Thus we see that there are two families of solutions in holographic model of d-wave superconductor which realize charge density wave.

Plots on Fig.\,\ref{Stripes} show the expectation values of the charged spin-2 operator as well: $\Delta_{\mu \nu} = \la O_{\mu \nu} \ra = \psi^{(1)}_{\mu \nu}$. 
In both types of the solutions the amplitude of the charge density modulation as well as the amplitude of oscillation of the order parameters grow when the wave vector of the given solution decreases. In both cases one can find a maximal value of the wave vector $k_{max}$ at which the amplitudes of all fields go to zero and the solution reduces to the empty Reisner-Nordstr\"om black hole (\ref{RN}). 

There are though some important differences between the two families of striped solutions.  
Firstly, as one can see on Fig.\ref{Stripes}, the amplitude of the $\Delta_{xy}$ order parameter in Type B is larger by the order of magnitude then that of the Type A. One can say that the type B striped solution is characterized by the oscillations of the $d_{xy}$ superconducting order parameter with zero mean value. These oscillations would lead to the breakdown of the long range superconducting order in the y-direction if the phase of the order parameter could be different in each wave, but this is not the case here. The modulus of the tensor VEV $|\Delta_{\mu \nu}|$ never vanishes along the wave profile, so the phase correlation remains present. One cannot associate though this behavior with any particular pattern of the gap in the density of states of fermionic excitations. Thus it is not obvious what kind of the features of superconductivity could this phase exhibit.

By comparing the behavior of $\Delta_{tx}$ component of the order parameter in two families of the solutions one can notice the following. While the amplitude of $\Delta_{tx}$ oscillation monotonically grows with decreasing $k_y$ in type A, in type B at the smallest $k_y$ it starts to decrease. Even though for the intermediate values of $k_y$ the $\Delta_{tx}$ amplitudes are of the same order in both types of the striped solutions, at smaller wave vectors $\Delta_{tx}$ becomes suppressed in type B solution and enhanced in type A. Thus we can associate the type A solution primarily with the oscillations of this component of the tensor order parameter.

\begin{figure}[ht]
\centering
\includegraphics[width=1\linewidth]{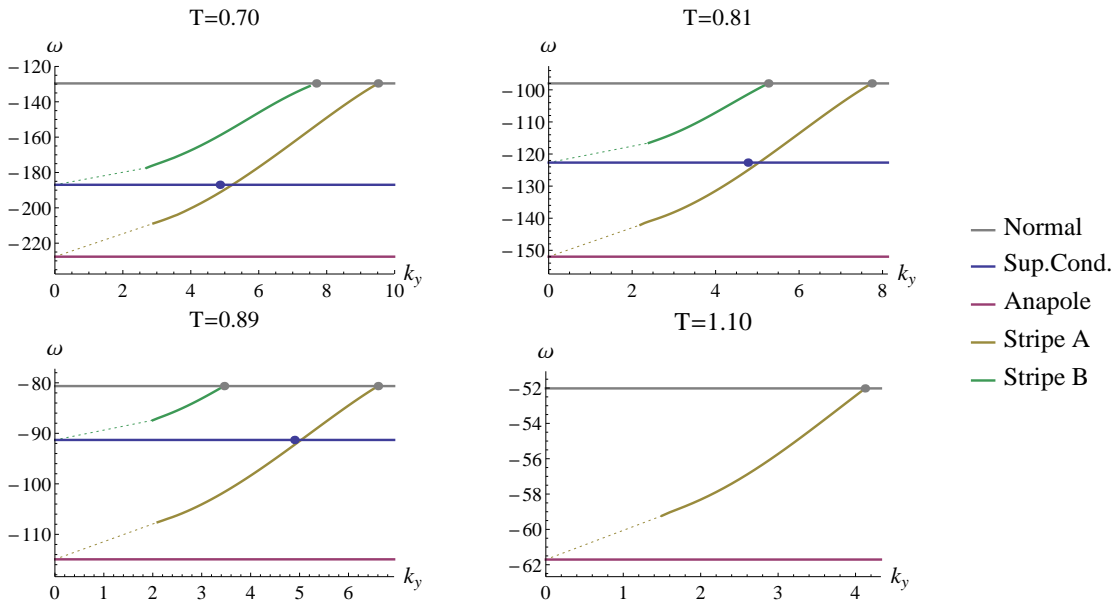} 
\caption{\label{Omega_plots} Mean thermodynamic potential of different solutions depending on the wave vector at various temperatures. The dimensional quantities are measured in terms of $z_h$, temperature is measured in terms of $T_c$. The thick dots represent the results of the perturbative study of instabilities in the homogeneous phases, see Fig.\,\ref{kcs} and Fig.\,\ref{NSks}. Dashed lines are eye-guides, we haven't got reliable numeric data in these regions. }
\end{figure}

The wavelength of the solution is an external parameter in our calculation procedure, i.e. the system is forced to have the given periodicity by putting it on a cylinder with compact $y$-direction. Thus in order to claim that the obtained solutions actually describe the stable state of the system we need to scan all possible wavelengths at constant temperature and study the mean thermodynamic potential of the obtained solutions as we are considering the system in the grand canonical ensemble. If $\omega(k_y)$ would have a minimum at certain $k^*_y$ at given temperature, this would mean that the stable striped phase with $k^*_y$ is spontaneously formed at this temperature even when the periodicity requirement is relaxed. The dependence of TD potentials of type A and type B solutions on the wave vector at various temperatures is shown on Fig.\ref{Omega_plots}. There exists the maximal value of the momentum $k_{max}$ beyond which, as we noted above, the amplitudes of the modulation of all the functions vanish and the striped solution smoothly connects with the Reisner-Nordstrom solution, which characterizes at this temperature the unstable normal state. When the momentum gets lower, TD potentials of both solutions monotonically decrease and there is no sign of the minimum at finite $k_y$. Given the fact that the problem under consideration is symmetric under $k_y \rar -k_y$ parity, it is suggestive to expect that the given striped phase will reach the minimum of the thermodynamic potential at $k_y=0$ while evolving to the certain homogeneous phase. 

This idea gets the support from the form of the striped solution at low $k_y$. On the second plots of Figs.\ref{Type_A} and \ref{Type_B} one can observe, that the value of the order parameter gets flattened around the points $\frac{1}{4}\lambda_y$ and $\frac{3}{4}\lambda_y$. At lower momenta this flattening becomes more pronounced while the curves around $\frac{1}{2}\lambda_y$ get steeper.\footnote{Because of this steepness we couldn't get precise numerical data at low $k$, as one can see on Fig.\ref{Omega_plots}.} This behavior means that while one increases the period of the solution, the wave profile decays into the homogeneous regions with finite constant values of $\psi_{tx}$ with opposite signs which are separated by the domain walls. The mean density of the TD potential in this configuration consists of the constant part associated with the homogeneous solution and the contribution from the domain walls. The latter drops inversely proportional to the wavelength $\lambda_y$. Given there are only 2 domain walls with constant potentials $\Omega_{d.w.}$ on one period, the spatial average value of $\omega$ will behave as $\omega_{d.w.} = 2 \Omega_{d.w.} / \lambda_y$ . Thus we expect that at $k_y \rar 0$ the thermodynamic potential of the striped solution will approach the one of the homogeneous phase. 

This behavior is seen for the type B phase on Fig.\ref{Omega_plots} which smoothly approaches the homogeneous $d_{xy}$ SC phase as $k_y \rar 0$. The thermodynamic potential of the type A phase, on the contrary, crosses the value of $\omega$ of the superconducting phase at finite wave vector and reduces further down. It is interesting to observe, that this crossing happens exactly at the value of critical momentum of SC phase calculated perturbatively in \cite{Krikun:2013iha} and shown on Fig.\,\ref{kcs}. Thick dots on Fig.\,\ref{Stripes} show the values of these critical momenta at given temperature and exhibit reasonable agreement between these two conceptually different calculations. The endpoint of the thermodynamic evolution of the type A striped phase is the new homogeneous phase which realizes the true minimum of TD potential and which apparently was not studied in the literature before. We will call it ``anapole'' phase as explained in Sec. \ref{Sec:anapole}.

Before we proceed further it is instructive to check our results by comparing the maximal values of the stripe wave vectors, which are observed in our solutions of the PDEs, with the momenta of the unstable modes in the  spectrum of the perturbative fluctuations of normal phase. We expect them to coincide because the striped phases reduce to the normal one exactly at these values of $k_y$.

\begin{figure}[ht]
\centering
\includegraphics[width=0.6 \linewidth]{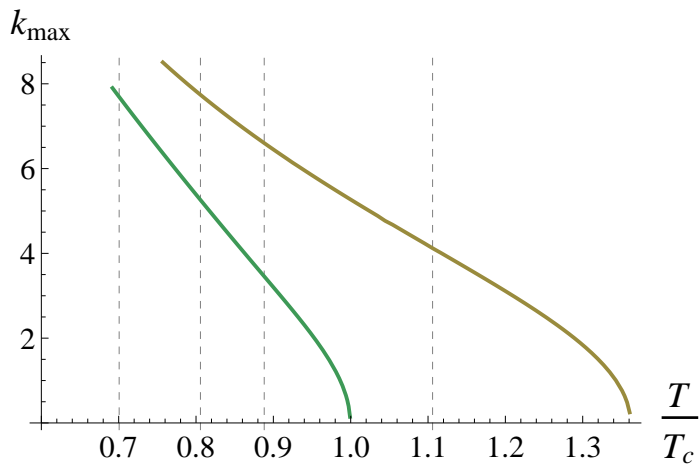} 
\caption{\label{NSks} The wave vectors of two unstable modes in the perturbative spectrum of the normal phase (measured in units of $z_h^{-1}$). The two curves are related to the two types of striped solutions. The grid lines show the temperatures used in plots of Fig. \ref{Omega_plots} }
\end{figure}

We study the linear perturbations of the Reisner-Norstr\"om solution with finite momentum $k$. We linearize the system of equations of motion (\ref{EOMs}) about the background (\ref{RN}) and look for the values of $k$, at which the nontrivial static solutions to the fluctuation equations with zero boundary conditions exist. The procedure is completely analogous to the one described in \cite{Krikun:2013iha}. In the end of the day, we find two branches of unstable fluctuations with different momenta which are shown on Fig.\ref{NSks}. One of the branches sets in exactly at $T=T_c$ and the critical values of $k$ at lower temperatures coincide exactly with the values of $k_{max}$ for the striped solution of type B, see Fig.\,\ref{Omega_plots}. This observation supports our argument that type B stripe is associated with fluctuating d-wave superconductor order parameter. The other branch sets in at higher temperature $T\approx 1.36 \, T_c$ and describes the maximum wave vectors of the stripes of type A. As the type A stripe is associated with the oscillations of $\Delta_{tx}$ and evolves to the anapole phase, we expect the critical temperature of the new phase to be equal to $1.36 T_c$.

\section{\label{Sec:anapole}Homogeneous anapole phase}
In this Section we are going to study the new homogeneous phase which arises as a result of  thermodynamic evolution of the Stripe A solution discussed above. It is suggestive to introduce $\psi_{tx}$ component of the tensor field as a part of the ansatz for this new solution. Indeed, one can check that the ansatz
\begin{equation}
A = A_t (z) dt, \qquad \psi_{t x} =  \psi_{t x}(z), \qquad \psi_{xz} = \psi_{z x}(z), \qquad \psi_{xy}=0
\end{equation}
provides a consistent truncation to the nonlinear equations of motion (\ref{EOMs}). The result is a system of 3 nonlinear ordinary differential equations. One should note that the equation $E_{tx}$ becomes algebraic and can serve to eliminate $\psi_{zx}$ from the problem.\footnote{It is worth noting here that, as it was in the striped case, the boundary parameters of the $\psi_{zx}$ field do not have a meaning of the source and VEV of any separate operator. They are completely defined by the boundary data of $\psi_{tx}$ which is dual to $\Delta_{tx}$ component of the order parameter.} In the end of the day we have only 2 second order ODEs which can be solved by means of the shooting method in complete analogy to the treatment of the $\psi_{xy}$ case in \cite{Benini:2010pr}. 

We obtain the boundary conditions at the horizon by power expanding the functions at $z=z_h$ and solving the equations of motion order by order. We find that the boundary data is defined by 2 constants
\begin{align}
\hat{\psi}_{tx}(z_h) &= \left. \frac{\psi_{tx}}{f(z)}\right|_{z\rar z_h} = C_1 + C_1 \left(  \frac{9 C_2^2}{8 m^2} - \frac{m^2}{6}\right) \frac{z - z_h}{z_h} + O\big((z-z_h)^2 \big), \\
\notag
\hat{A}_t (z_h) &= \left. \frac{A_{t}}{f(z)}\right|_{z\rar z_h} = C_2 - C_2 \left( 1 + \frac{9 C_1^2}{8 m^2} \right) \frac{z - z_h}{z_h} + O\big((z-z_h)^2 \big).
\end{align}
For a given value of temperature, which defines $z_h$ via (\ref{metric}) we use a shooting method to find $(C_1,C_2)$ which would lead to the appropriate boundary conditions at the AdS boundary $z=0$
\begin{equation}
 A_{t}(0) = \mu, \qquad \psi_{tx}(0)=0.
\end{equation}

\begin{figure}[ht]
\centering
\includegraphics[width=0.6 \linewidth]{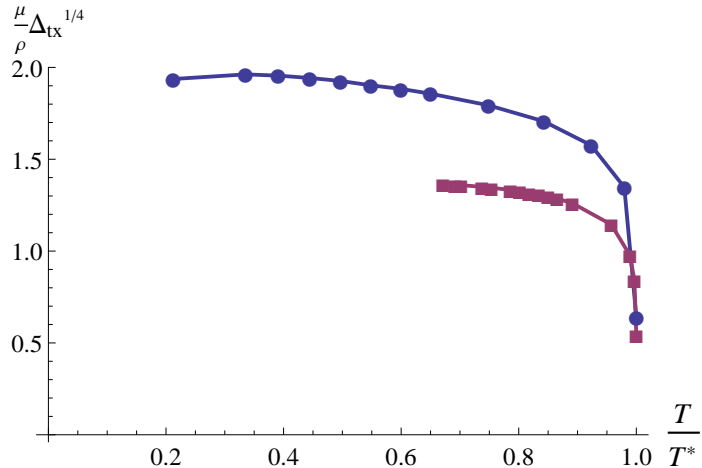} 
\caption{\label{anapole_condensates} The value of $\Delta_{tx}$ order parameter depending on the temperature in anapole phase for $m^2=4$ (upper curve) and $m^2=1.75$ (lower curve). The temperature is measured with respect to the critical temperature $T^*$ when the anapole phase sets in. The normalization of $\Delta_{tx}$ is chosen in analogy with the plots of \cite{Benini:2010pr}}
\end{figure}

Similarly to the d-wave superconducting case we find that below certain temperature $T=T^*$ the nontrivial solution to the equations of motion exist and describes the phase with nonzero value of $\Delta_{tx}$ order parameter, see Fig.\,\ref{anapole_condensates}. It is interesting to study the physical features of this order parameter. From the nonrelativistic point of view the time component of the tensor is seen as a vector (i.e. the time component of the energy-momentum tensor is a vector of momentum). Thus the phase, which we are describing, is characterized by a polar in-plane vector order parameter which is also odd under the time reversal transformation
\begin{equation}
 P(\Delta_{tx}) = - \Delta_{tx}, \qquad  T(\Delta_{tx}) = - \Delta_{tx}.
\end{equation}
These features are very different from the d-wave superconducting order parameter which is parity and time-reversal even. In condensed matter literature one can find a discussion of similar order parameter which is claimed to be present in the pseudogap phase of cuprates \cite{varma2006theory} and it is called ``anapole moment''. This is a reason why we adopt this name for the new homogeneous phase of holographic model of d-wave superconductor. There is though one important difference. Our order parameter is complex and breaks $U(1)$ symmetry similarly to the superconducting order parameter. Contrary to that there are no indications that the anapole moment is charged and brakes $U(1)$ electromagnetic group as well.

\begin{figure}[ht]
\centering
\includegraphics[width=0.6 \linewidth]{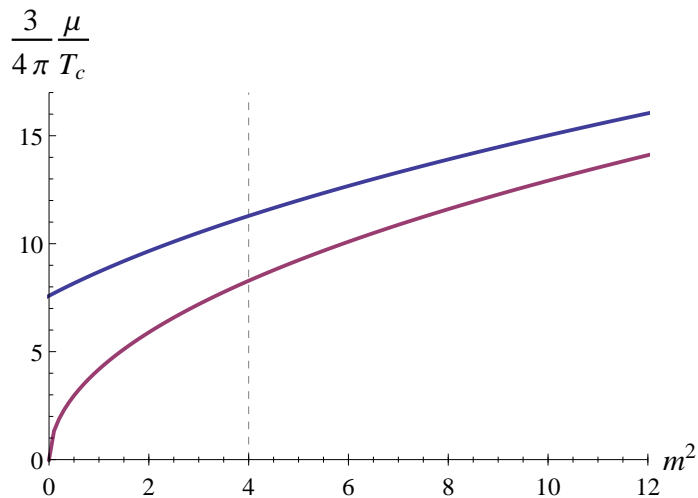} 
\caption{\label{Tcs} The inverse transition temperatures of the two homogeneous phases: d-wave superconductor and anapole. The critical temperature of the anapole phase is higher for all allowed values of $m^2$. The grid line shows $m^2=4$, used for calculations in Sec.\,\ref{stripes}}
\end{figure}

As we observed in the previous Section, the critical temperature of the new phase for $m^2=4$ is larger then that of the SC phase. It is interesting to study whether this is always the case at any values of $m^2$. One can study the critical temperatures in a perturbative approach using the fact that the values of the tensor field are small near the transition point \cite{Kim:2013oba}. Taking this fact into account we reduce the problem to the solution of linear ODE and find the critical temperatures of the anapole phase at various values of $m^2$ as it is shown on Fig.\,\ref{Tcs} (Note that it is inverse temperature which is plotted here). The transition temperature of the new phase is always higher then that of the d-wave superconducting phase. Curiously enough it goes to infinity when $m^2=0$. One can explain it by noting that at $m^2 =0$ the gauge symmetry of the tensor field is restored and the mode which constitutes the anapole phase reduces to the gauge translation.

\begin{figure}[ht]
\centering
\includegraphics[width=0.6 \linewidth]{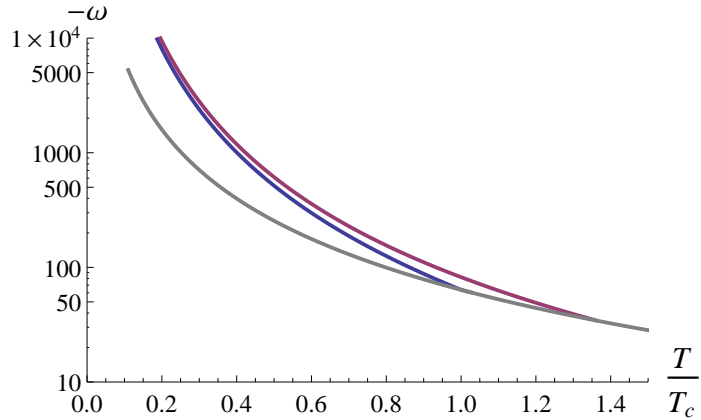} 
\caption{\label{OmsM} The thermodynamic potential density of the homogeneous phases: anapole, d-wave superconductor and normal (counting from top) depending on the temperature for $m^2=4$. The anapole phase is always thermodynamically preferred.}
\end{figure}

If the anapole phase has higher transition temperature, one can expect that it will remain thermodynamically preferred even when the temperature is lowered. Nonetheless the TD potential of the d-wave phase can decrease fast enough to make the SC phase preferred at some low enough temperature. We check this possibility by calculating the TD potentials of anapole and d-wave phases in the range of temperatures below transition. For $m^2=4$, as one can see  on Fig.\,\ref{OmsM}, anapole phase remains preferred even for low $T$ and the d-wave phase just tries to catch up. Similar behavior is seen for $m^2=1.75, 10$ and $16$ and there are no signs for any conceptual change in it at lower or higher masses. So we have to conclude that the d-wave superconducting phase is never stable in the model under consideration.

\section{\label{conclusion}Conclusion}

In this work we found several new solutions to the equations of motion of the holographic model of d-wave superconductor. It is interesting to compare the basic symmetry properties of the phases of the dual theory associated with these solutions. Firstly recall that the d-wave superconducting phase with order parameter $\Delta_{xy}$ studied in \cite{Benini:2010pr} is time reversal and parity even and homogeneous. The complex order parameter breaks the rotation symmetry to the discrete set of rotations on $\pi/2$ and breaks global $U(1)$ symmetry leading to the superfluid (superconducting) phase.

The striped phases, which we study in Section \ref{stripes}, break the translational symmetry spontaneously. Each solution describes the phase which is modulated only in one spatial direction, but one should recall that because the Lagrangian is symmetric upon rotation interchange of the coordinates $x\lr y$, the stripes in two transverse directions (i.e. along $x$- or $y$-axis) are completely degenerate. Thus in generic configuration one should expect to find different domains with the different orientation of the stripes. The mean properties of such configuration would have discrete symmetry under rotations on $\pi/2$, even though each given domain is symmetric only under rotation on $\pi$.

The nonzero values of the component $\Delta_{xt}$ in the striped phases break $P$- and $T$-symmetry locally. But one should note that the discrete set of translations remains the symmetry of the stripes. The action of the parity or time-reversal transformation is equivalent to the overall shift of the stripe on half-period (combined with the gauge transformation in case of T). I.e. the striped state is symmetric under the combination of $P$ or $T$ and half-period shift. Hence its response to any macroscopic (insensitive to the shifts of the order of the wavelength) probe will still be time-reversal and parity even. 

As it was already mentioned in Section \ref{stripes}, in the striped phases there are no points were the modulus of the tensor order parameter would vanish. Thus the phase of the tensor components $\Delta_{xy}$ and $\Delta_{tx}$ remains correlated along the whole wave profile. This suggests that such a phase would support at least some features of superconductivity. On the other hand, the gap in the quasiparticle spectrum should be proportional to the VEV of the tensor field. As the VEVs of different tensor components oscillate spatially with zero mean value, it is not clear whether any gap can be observed in the spectrum and this question certainly deserves further study.  

In contrast to the striped phases the anapole phase preserves translations and the mean value of the order parameter $\Delta_{tx}$ is nonzero. Thus parity and time-reversal symmetries are spontaneously broken even on the macroscopic scales. The in-plane anapole moment picks a certain direction breaking the rotations symmetry. But similarly to the case with the stripes, the phases with two perpendicular directions are degenerate, so in the anapole phase one should expect to observe a number of the domains with different orientation of the anapole moment with the overall $\pi/2$ rotational symmetry.   

One important difference between the anapole moment observed in this work and anapole moment, or loop current order, claimed to be present in cuprates \cite{varma2010high} is that the former is complex. Condensation of this order parameter breaks the global $U(1)$  group similarly to the superconducting order. On the other hand  the loop current order, if it can be understood as the specific molecular current inside the unit cell, is not expected to be complex and should preserve $U(1)$.

Our thermodynamic treatment shows that in the present model the anapole phase has the lowest thermodynamic potential among all studied solutions at all values of the model parameters. 
The two types of charge density wave solutions describe the transient phases  which interpolate between the normal phase and two types of the homogeneous condensed phases. The d-wave SC phase possess dynamical instability which drives the transition to the anapole phase. Even though this situation is phenomenologically unsatisfactory, we find it very encouraging that the simple model that we use encompass this wide range of solutions just due to the nontrivial kinematics of the charged massive symmetric tensor field. It would be interesting to study possible modifications of the model (i.e. by expanding the gauge field sector) and observe whether the competition between phases can be induced by additional interaction and be driven by some parameters of the system. We should also note here that in  more realistic models the interaction with the underlying crystal lattice could promote the striped phases to be thermodynamically stable and that would make the phenomenology much more interesting. In this case it would be natural to expect the charge density waves to be commensurate with the lattice. 

We find that generically at large wavelengths the given striped phase decays into the domains of the corresponding homogeneous phase separated by the domain walls. One can expect such domain walls to be formed in the nonequilibrium phase transition processes by the Kibble-Zurek mechanism. So even though the striped phases, that we observed here, are not thermodynamically stable, they can leave an imprint on the ground state by being involved in the transition processes. In view of this fact it would be especially interesting to study the features of the domain walls which are produced this way.

\acknowledgments
I am grateful to Aristomenis Donos and Sergey Pershoguba for valuable comments and to Gerome Gauntlett, Marika Taylor, Kostas Skenderis and Konstantin Zarembo for fruitful discussions. 
Author acknowledges the hospitality of Imperial College London and Southampton University where the preliminary results of this project were discussed. I would also like to thank the organizers of ``International Workshop on Condensed Matter Physics \& AdS/CFT'' in Kavli IPMU for a hospitality and an opportunity to present this work.

This work is partially supported by RFBR grant 15-02-02092 and Dynasty Foundation.

 \appendix

 \section{\label{App:Numerics} Numerical setting for the calculation of the striped phase}
 
Before proceeding to the numerical calculations we rescale the coordinates and the functions in order to work with dimensionless quantities. The coordinates are rescaled as
\begin{equation}
z \rar z_h \, \tilde{z} , \qquad y \rar  \lambda_y \,\tilde{y} \equiv \hat{\lambda} z_h \,\tilde{y}.  
\end{equation}
After this redefinition the domain of computation reduces to the unit square. By redefining the functions we aim to achieve two goals: scale out the charge $q$ and get simple Dirichlet boundary conditions at the AdS boundary and finite values on the horizon. These are achieved by the rescaling
\begin{align}
\label{rescale}
\psi_{xy} &\rar q^{-1} \tilde{z}^{ (3 - \sqrt{9+4 m^2})/2} \tilde{\psi}_{xy}, \\
\notag
\psi_{tx} &\rar q^{-1} \tilde{z}^{ (3 - \sqrt{9+4 m^2})/2} f(\tilde{z}) \tilde{\psi}_{tx}, \\
\notag
\psi_{xz} &\rar q^{-1} \tilde{z}^{ (5 - \sqrt{9+4 m^2})/2} \tilde{\psi}_{xz}, \\
\notag
A_t &\rar \frac{\mu}{q z_h} f(\tilde{z}) \tilde{A}_t \equiv \hat{\mu} f(\tilde{z}) \tilde{A}_t.
\end{align}
With the above redefinitions the boundary conditions (\ref{bcB}) reduce to the Dirichlet type
\begin{equation}
\tilde{\psi}_{xy}(0,y) =\tilde{\psi}_{tx}(0,y) =\tilde{\psi}_{zx}(0,y) = 0, \qquad  \tilde{A}_t(0,y)=1
\end{equation}
and regularity on the horizon means
\begin{equation}
\tilde{\psi}_{xy}(1,y), \ \tilde{\psi}_{tx}(1,y), \ \tilde{\psi}_{zx}(1,y), \ \tilde{A}_t(1,y) - \mbox{finite.}
\end{equation}
By expanding the equations of motion (\ref{EOMs}) near the horizon and making use of the constraint (\ref{Constr:striped}) we get the following system of nonsingular differential equations on the fields and their first $z-$derivatives
\begin{gather}
 \label{boundary_eq}
 \frac{1}{\hat{\lambda}^2}\p_{\tilde{y}}^2 \tilde{\psi}_{xy} - 3 \p_{\tilde{z}} \tilde{\psi}_{xy} - \frac{1}{2}\left(9 + 2 m^2 - 3 \sqrt{9+4 m^2} \right) \tilde{\psi}_{xy} - \frac{27 \hat{\mu}}{2 m^2 \hat{\lambda} } \p_{\tilde{y}}(\tilde{\psi}_{tx} \tilde{A}_t) = 0,\\
 \notag
  \frac{1}{\hat{\lambda}^2}\p_{\tilde{y}}^2 \tilde{\psi}_{tx}-6  \p_{\tilde{z}} \tilde{\psi}_{tx} - \left(9 + m^2 - 3 \sqrt{9+4 m^2} \right) \tilde{\psi}_{tx} + \frac{\hat{\mu}}{2 \hat{\lambda}} \p_{\tilde{y}}(\tilde{\psi}_{xy} \tilde{A}_t) - \frac{3}{2} \hat{\mu} \tilde{\psi}_{xz} \tilde{A}_t =0, \\
 \notag
 \frac{1}{\hat{\lambda}^2}\p_{\tilde{y}}^2 \tilde{\psi}_{xz} - m^2 \tilde{\psi}_{xz} - \frac{1}{2}\left(3 - \sqrt{9+4 m^2} \right) \frac{\p_{\tilde{y}}\tilde{\psi}_{xy}}{\hat{\lambda}} - \frac{\p_{\tilde{y}} \p_{\tilde{z}}\tilde{\psi}_{xy}}{\hat{\lambda}} - \frac{9}{2} \hat{\mu} \tilde{\psi}_{tx} \tilde{A}_t = 0, \\
 \notag
 \frac{1}{\hat{\lambda}^2}\p_{\tilde{y}}^2 \tilde{A}_{t} - 6 \p_{\tilde{z}} \tilde{A}_{t} - (6 + \tilde{\psi}_{xy}^2) \tilde{A}_{t} + \frac{1}{2 \hat{\lambda} \hat{\mu}} \left(\tilde{\psi}_{tx} \p_{\tilde{y}} \tilde{\psi}_{xy}  - \tilde{\psi}_{xy} \p_{\tilde{y}} \tilde{\psi}_{tx} \right) + \frac{3}{2 \hat{\mu}} \tilde{\psi}_{tx} \tilde{\psi}_{xz} = 0.
\end{gather}
In the end of the day there are three dimensionless parameters: $m^2$, $\hat{\mu} = \mu z_h$ and $\hat{\lambda} = \lambda_y z_h = \frac{2 \pi z_h}{k_y}$, which enter the problem. And we get the solutions for different choices of all three.

We construct the numerical procedure by approximating derivatives with finite differences. We use pseudospectral approximation for the derivatives what means that we are forced to use Chebyshev grid in $z$ dimension and equal spacing grid in $y$-dimension, because $y$-dimension is periodic \cite{trefethen2000spectral}. The grid covers the boundaries of the integration domain, because the boundary conditions that we use are regular. The grid of the  size $20_y \times 25_z$ proved to be sufficient for the study of general behavior and free energies of the striped solutions . 

We make use of \texttt{NDSolve'FiniteDifferenceDerivative} function in \textit{Wolfram Mathematica 9} \cite{Mathematica9} in order to obtain the differentiation matrices. At each step of the Newton-Raphson procedure we linearize the equations of motion and construct a matrix of the linear differential operator which acts on the function increments. Being the functional of the current field values this operator must be recalculated after each step. The resulting linear algebraic system is solved be means of \texttt{LinearSolve} function. At the final stages of convergence the operation of inverting the matrix of the linear operator start to introduce significant numeric errors, so it proved to be more efficient to use \texttt{LeastSquares} function instead of \texttt{LinearSolve}, which minimizes the mean square value of the equations of motion at the nodes. Finally, we repeat the Newton-Raphson procedure until the largest value of the equations of motion becomes numerically small ($\sim 10^{-9}$) and check whether the constraint (\ref{Constr:striped}) is satisfied.

In order to obtain the boundary data we interpolate the discretized functions with maximal interpolation order and apply then compare the interpolated functions with (\ref{bcB}).

\bibliographystyle{unsrt}
\bibliography{checkerboard2.bib}

 \end{document}